# Compact engineering of path-entangled sources from a monolithic quadratic nonlinear photonic crystal

H. Jin[1], P. Xu[1]*, X. W. Luo[1], H. Y. Leng[1], Y. X. Gong[2] and S. N. Zhu[1]

[1]National Laboratory of Solid State Microstructures and College of Physics, Nanjing University, Nanjing 210093, China.

[2] Department of physics, Southeast University, Nanjing 211189, China.

* Author to whom correspondence should be addressed. Electronic mail: pingxu520@nju.edu.cn).

**Abstract**

Photonic entangled states lie at the heart of quantum science for the demonstrations of quantum mechanics foundations and supply as a key resource for approaching various quantum technologies. An integrated realization of such states will certainly guarantee a high-degree of entanglement and improve the performance like portability, stability and miniaturization, hence becomes an inevitable tendency towards the integrated quantum optics. Here, we report the compact realization of steerable photonic path-entangled states from a monolithic quadratic nonlinear photonic crystal. The crystal acts as an inherent beam splitter to distribute photons into coherent spatial modes, producing the heralded single-photon even appealing beamlike two-photon path-entanglement, wherein the entanglement is characterized by quantum spatial beatings. Such multifunctional entangled source can be further extended to high-dimensional fashion and multi-photon level as well as involved with other degrees of freedom, which paves a desirable way to engineer miniaturized quantum light source.

Photons especially entangled photons are an appealing information carrier for a lot of available degrees of freedom and their inherent advantages like low-loss, high-speed transmission and robust coherence, hence photonics has been a leading approach in realizing the future quantum technologies[1]. Nowadays a consequent tendency towards the practical quantum information processing is to manipulate the photons on a mono-platform like an integrated waveguide chip[2-12]. However, most of the optical chips require an external quantum light source, thereby, to move a step further towards integrated quantum optics, developing internal quantum light source inside the chip or miniaturizing the external light source deserves deep and sustaining investigations.

A solid strategy for achieving integrated multifunctional quantum light source turns to the traditional nonlinear optical crystals especially the domain-engineered quadratic nonlinear photonic crystals (NPC)[13,14]. By domain-engineering technique, the spatial and temporal properties of entangled photons can be controlled inherently during the quasi-phase-matching (QPM) spontaneous parametric downconversion (SPDC) processes[15-23], and the polarization-entangled, frequency-entangled states etc. can be directly obtained[24-28]. In this work, we mainly concentrate on the integrated engineering of single- and two-photon path-entangled states. The path-entanglement is a typical entangled source which applies the spatial modes to encode information, and has been harnessed for a variety of applications, such as quantum precise phase-measurement[29,30] and super-resolution quantum lithography[31,32]. Even for the simple and economic single-photon path-entanglement[33-36], people also find interesting applications in quantum teleportation[37] and quantum network[38]. However, for achieving path-encoded qubits, extra endeavors should be paid. For example, a two-photon NOON state, which can be usually generated by Hong-Ou-Mandel interference[2,39,40], requires a beam splitter to cascade after the nonlinear crystal and hold on at a balanced position, therefore integrated realization of such photon source is of essential importance.

Here, in this work we report the experimental demonstrations of the direct generation of single- and two-photon path-entangled sources from a monolithic domain-engineered nonlinear photonic crystal (NPC). Resulting from the concurrent multiple QPM SPDC processes, versatile spatial forms of down-converted beams are achieved and then various photonic path-entangled states can be generated and transformed inside the same crystal wafer. Specifically, the heralded single-photon path-entanglement can be steered into beamlike two-photon path-entanglement. The path-entanglement is physically guaranteed by the coherence between concurrent nonlinear processes. When combined with other functions by the design of microstructure, such monolithic domain-engineered crystal presents unique advantages in engineering novel quantum light sources which is compact and miniaturized, hence geared to the integrated quantum optic.

Experimentally we design and fabricate a hexagonally poled lithium tantalate (HPLT) crystal for the engineering of integrated path-entangled photon source. Figure 1(a) shows domain structure of the etched HPLT sample which is qualified for the efficient generation of degenerate 1064 nm photon pairs. The quadratic nonlinear coefficient of our crystal can be expressed as Fourier series of $\chi^{(2)}(\vec{r}) = d_{33} \sum_{m,n} f_{m,n} e^{i\vec{G}_{m,n}\cdot\vec{r}}$. The corresponding reciprocal lattice sketched in Fig. 1(b) shows a six-fold symmetry which can supply two or more reciprocal vectors $\vec{G}_{m,n}$ (m, n are integers) for fulfilling the QPM condition $\vec{k}_p - \vec{G}_{m,n} - \vec{k}_s - \vec{k}_i = 0$ and ensuring the multiple geometries for the entangled photons generation. In this work,

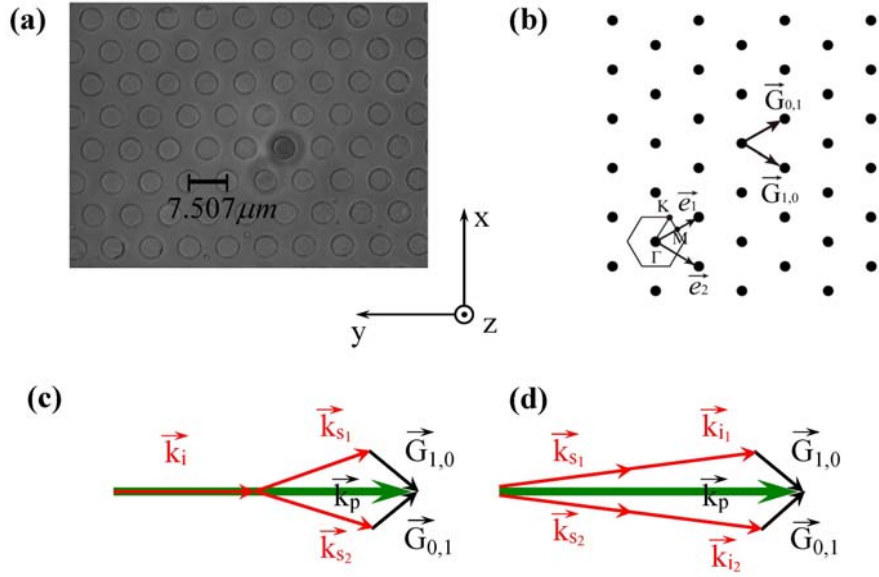

**FIG. 1 (a)** Micrograph of the etched congruent HPLT with $a$=7.507 μm and a reversal factor of $r/a$.~28% ($r$ is the radius of round domain-inverted areas). The crystal is a thick wafer with dimensions of 18 mm×6mm×0.5 mm. The polarization configuration of concerned QPM process is $e \to e+e$ and the maximum nonlinear coefficient $d_{33}$ is involved. **(b)** The reciprocal lattice of the crystal. $\vec{G}_{1,0}$ and $\vec{G}_{0,1}$ are two reciprocal vectors which are used to support two concurrent SPDC processes. **(c)** Phase matching condition for the generation of single-photon entanglement. When two concurrent SPDC processes share the same wave vector of idler $\vec{k}_i$, the signal photon can be emitted in either mode 1 or mode 2 coherently. **(d)** Phase matching condition for the generation of two-photon entanglement. A beam-like photon pair can be emitted in either mode 1 or mode 2 coherently.

we mainly focus on the QPM SPDC processes fulfilled by reciprocal vectors $\vec{G}_{1,0}$ and $\vec{G}_{0,1}$, which have equal Fourier coefficients $f_{1,0} = f_{0,1} = 0.32$ indicating an effective nonlinearity of $0.32 d_{33}$.

When the crystal is pumped by a 532 nm laser along y-axis, two concurrent SPDC processes will be ensured under a pair of noncollinear QPM geometries following $\vec{k}_p - \vec{G}_{1,0} - \vec{k}_s - \vec{k}_i = 0$ and $\vec{k}_p - \vec{G}_{0,1} - \vec{k}_s - \vec{k}_i = 0$, wherein $\vec{k}_p$, $\vec{k}_s$ and $\vec{k}_i$ are the wave vectors of the pump, signal and idler photons, respectively. Generally, the photon pair will emit as either one of conical beams with principle axis along $\vec{k}_p - \vec{G}_{1,0}$ or $\vec{k}_p - \vec{G}_{0,1}$. Since we only pay attention to the spectral part we can write the two-photon state under the first-order perturbation approximation[41],

$$|\psi\rangle = \Psi_0 \int d\omega_s \int d\omega_i \phi(\omega_s, \omega_i) \left[ \hat{a}_{s_1}^\dagger(\vec{k}_{s_1}) \hat{a}_{i_1}^\dagger(\vec{k}_{i_1}) + \hat{a}_{s_2}^\dagger(\vec{k}_{s_2}) \hat{a}_{i_2}^\dagger(\vec{k}_{i_2}) \right] |0\rangle, \quad (1)$$

in which $\Psi_0$ is a normalization constant, and the subscripts represent the signal (s) and idler (i) in mode 1 and mode 2, respectively. The two-photon mode functions in the two processes take the same form $\phi(\omega_s, \omega_i) = \text{sinc}(\Delta k_y L / 2) \delta(\omega_p - \omega_s - \omega_i)$, wherein $L$ is the length of the crystal, $\Delta k_y$ is the phase mismatching in the longitudinal direction, and $\omega_p$, $\omega_s$, $\omega_i$ are angular frequencies of pump, signal and idler, respectively. Here only two types of concurrent QPM SPDC processes are of our interest, which are depicted in Figs. 1(c) and 1(d), indicating the direct generation of single-photon and two-photon path-entangled states, respectively. In Fig. 1(c), two SPDC processes share one photon, namely the idler, which propagates collinearly with the pump, thereby, the corresponding signal photon belongs to either mode 1 ($s_1$) or mode 2 ($s_2$). Following the previous paper[42], we can take a partial trace of $\hat{\rho} = |\psi\rangle\langle\psi|$, which is the density matrix operator of two-photon state, and write the density matrix of signal as

$$\hat{\rho}_s = \Psi_0^2 \int d\upsilon |\phi(\upsilon)|^2 \left[ a_{s_1}^\dagger(\frac{\omega_p}{2} + \upsilon) + a_{s_2}^\dagger(\frac{\omega_p}{2} + \upsilon) \right] |0\rangle\langle 0| \left[ a_{s_1}(\frac{\omega_p}{2} + \upsilon) + a_{s_2}(\frac{\omega_p}{2} + \upsilon) \right], \quad (2)$$

in which we have taken advantage of the $\delta$-function and introduced a detuning frequency $\upsilon$. Now the spectral mode function is expressed as $\phi(\upsilon) = \text{sinc}(\Delta k_y L / 2)$,

in which the phase mismatching in the longitudinal direction is

$$\Delta k_y = \frac{\upsilon(1-\cos\theta_1)}{u_s} - \frac{\upsilon^2(1+\cos\theta_1)}{2}\frac{d}{d\omega}\left(\frac{1}{u}\right)\bigg|_{\omega=\omega_p/2}, \text{ wherein } u_s = \frac{d\omega}{dk}\bigg|_{\omega=\omega_p/2} \text{ is group}$$

velocity of signal and $\theta_1$ is the angle between the wave vectors of signal and pump in the crystal. The bandwidth (FWHM) of signal photons is calculated to be about 35.4 nm. When using single-frequency approximation which can be guaranteed by narrowband interference filters, we can simplify Eq. (2) to

$$|\psi_1\rangle = \frac{1}{\sqrt{2}}(|1,0\rangle + |0,1\rangle), \qquad (3)$$

which is a single-photon path-entangled state. Figure 1(d) shows another type of two concurrent SPDC processes supported by $\vec{G}_{1,0}$ and $\vec{G}_{0,1}$ when increasing the temperature of the HPLT crystal. Notably, in this case the photon pair emit together into beamlike modes which are desirable for the high-efficiency collection, and we can deduce the two-photon state to be

$$|\psi\rangle = \Psi_0 \int d\upsilon \phi(\upsilon)\left[\hat{a}_{s_1}^\dagger(\frac{\omega_p}{2}+\upsilon)\hat{a}_{i_1}^\dagger(\frac{\omega_p}{2}-\upsilon) + \hat{a}_{s_2}^\dagger(\frac{\omega_p}{2}+\upsilon)\hat{a}_{i_2}^\dagger(\frac{\omega_p}{2}-\upsilon)\right]|0\rangle. \qquad (4)$$

In this case the phase mismatching is $\Delta k_y = -\upsilon^2\cos\theta_2\frac{d}{d\omega}\left(\frac{1}{u}\right)\bigg|_{\omega=\omega_p/2}$, wherein $\theta_2$ is the angle between the wave vectors of signal and pump. The theoretical bandwidth (FWHM) of down-converted photons is about 31.4 nm. When using single-frequency approximation, we can simplify Eq. (4) to

$$|\psi_2\rangle = \frac{1}{\sqrt{2}}(|2,0\rangle + |0,2\rangle), \qquad (5)$$

which is a two-photon NOON state.

Figures 2(a)-2(d) record the spatial distribution evolution of down-converted photons generated from the HPLT when the temperature increased. Figures 2(b) and

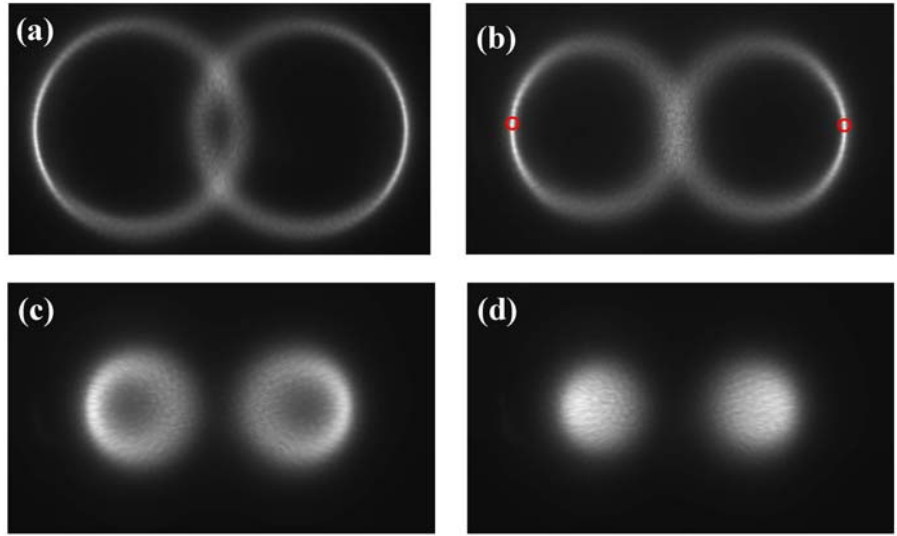

**FIG. 2 (a)** The spatial Fourier spectrum obtained at the focal plane of a convex lens when the crystal is 145 °C by a sensitive CCD camera. **(b)** The spatial Fourier spectrum at 154.6 °C. The two red circles indicate the two modes of the single-photon entangled state. The angle between two signal modes outside of the crystal is measured to be about 9.56° which consists well with theoretical design of 9.38°. **(c)** The spatial Fourier spectrum at 168 °C **(d)** The spatial Fourier spectrum at 172.3 °C. The two bright spots correspond to the noncollinear beam-like modes of two-photon entangled state. The emitting angle between the two modes outside of the crystal is about 4.81°, while the theoretical design is 4.68 °, and the measured divergence angle of each mode is about 2.98°.

2(d) disclose the spatial distribution of single-photon and two-photon path-entanglement sources when the QPM conditions of Figs. 1(c) and 1(d) are satisfied, respectively. Although the coherence between two single-photon or two-photon modes is guaranteed physically by the concurrent coherent nonlinear processes sharing the same pump, it is necessary to verify the path-entanglement experimentally, therefore we carry out the single-photon and two-photon spatial beating experiments.

At 154.6 °C, as shown in Fig. 2(b) the photon pair emit into either one of two tangent cones, resulting in the heralded single-photon path-entanglement. When the idler photon is found to locate in the overlapped region of two cones, the corresponding signal photon will emit into two possible modes denoted by red circles. Theoretically, when two coherent signal modes with wavelength $\lambda$ intersect at an angle of $2\theta$ (incidence angle $\theta$ each), the spatial beating fringes will appear as[32]

$$I_{s.c.} \propto 1 + \cos\left(\frac{2\pi x}{\lambda/(2\sin\theta)}\right). \qquad (6)$$

Figure 3(a) shows the experimental setup for observing single-photon spatial beating. The coherent signal modes are picked out by two slits and intersect at an angle of $2\theta$=10.4 mrad. Their beating fringes are recorded by a CCD camera which is sensitive to a few photons. In Fig. 3(b) we show the interference fringe when two optical paths are precisely controlled to be equal. Figure 3(c) reveals its intensity distribution. The experimental data (blue dots) is fitted well with a sinusoidal curve weighed by a Gaussian profile. The fitted curve shows a period of 104.4 μm, which agrees with the calculated value. The fringe visibility is 70.9%, which confirms that the two modes sketched in Fig. 1(c) are generated in a coherent way. The coherence is due to the indistinguishability of their idler modes. The coherence length, which is estimated by recording the relationship between the fringe visibility and the path difference as shown in Fig. 3(d), is measured to be 38.2 μm, which fits well with theoretical value of 39.9 μm deduced from the theoretical single-photon bandwidth after taking the interference filter's transmission profile into account. When tuning the temperature away from 154.6 ºC, two signal modes will share less of the idler mode and tend to be incoherent. At 143 ºC, we find that the fringe visibility decreases to less than 30% even with balanced path lengths of two modes.

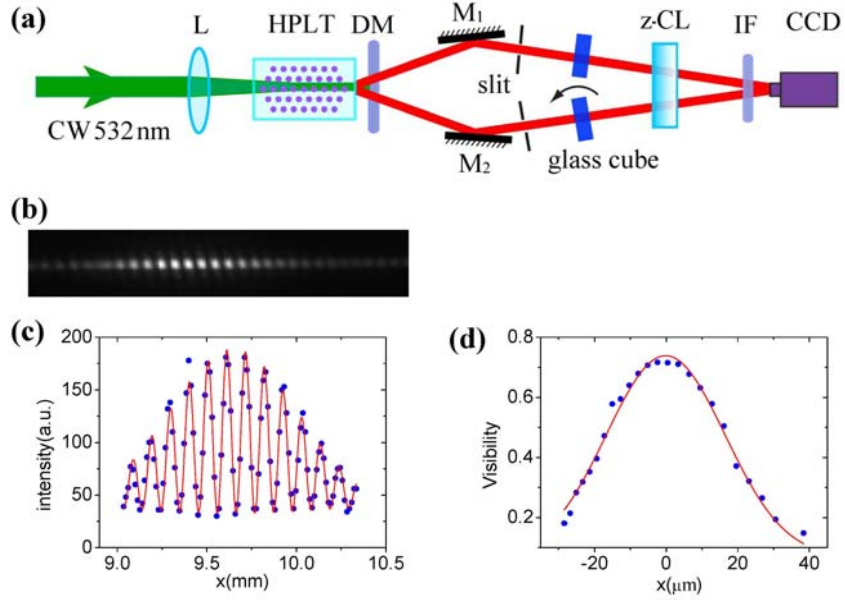

**FIG. 3 (a)** Experimental setup. The crystal is controlled at 154.6 ºC and pumped with a CW single longitudinal mode 532 nm laser. After the crystal, the down-converted photons of 1064 nm are first separated from the pump by a dichromatic mirror DM, then two signal modes are both reflected by flat mirrors to ensure an small intersecting angle $2\theta$=10.4 mrad so that the beating fringes can be resolved by the CCD camera which is sensitive to a few photons. A cylindrical lens z-CL gathers the intensity along z-axis. For each path, a 500 μm width slit (corresponding to an arc of 43 mrad on the ring of down-converted photons) is used to pick out the coherent signal mode. Preceding the CCD is a 40 nm bandwidth interference filter centered at 1064 nm. The optical path length of each mode is adjusted by rotating a K9 glass cube with 25.4 mm thickness. **(b)** The interference fringe observed by the CCD when the two modes have no optical path difference. **(c)** The intensity distribution extracted from Fig. 2**(b)**. The solid red curve is sinusoidal fitting weighed by a Gaussian profile. **(d)** The relationship between the fringe visibility and the optical path-length difference.

As shown in Fig. 2(d), when the crystal is controlled at 172.3 ºC, the photon pair will emit into either one of the beamlike modes, resulting in the two-photon path-entanglement. The photon pair are entangled over two modes denoted by the bright well-defined beamlike spots. Theoretically, the two-photon spatial correlation is

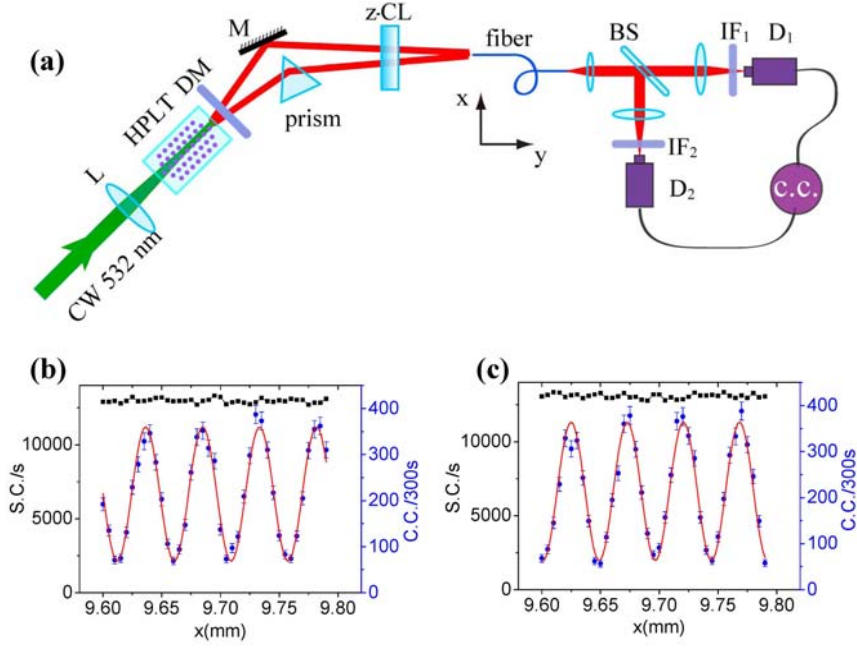

**FIG. 4 (a)** Experimental setup. The temperature of the crystal is controlled at 172.3 °C. A prism and a reflecting flat mirror are used to ensure that the two modes intersect at a small angle, and a cylindrical lens z-CL gathers the photons along the z-axis to enhance the coincidence counts collected by the fiber. An optical fiber is used to scan along the x-axis and cascaded by a two-photon coincidence counting measurement. Specifically, the entangled photons are separated by a beam splitter (BS) after collimated by a lens and collected into two single photon detectors $D_1$ and $D_2$, respectively, by two lenses. The bandwidth of the interference filter placed before each detector is 10 nm. **(b)** Measured single (black squares) and coincidence (blue dots) counts versus the transverse position of the fiber tip. The red solid line is sinusoidal fitting. **(c)** Measured single (black squares) and coincidence (blue dots) counts when one of the optical path lengths is extended by about 1 mm. Error bars shows $\pm\sqrt{R_{c.c}}$.

proportional to[32]

$$R_{c.c.} \propto 1 + \cos\left(\frac{2\pi x}{\lambda/(4\sin\theta)}\right) \qquad (7)$$

when two two-photon modes intersect at an angle of $2\theta$. Two-photon beating fringes should present the period half of the single-photon case with the same wavelength.

Figure 4(a) is the experimental setup for observing two-photon spatial beating fringes. A prism and a reflecting flat mirror are used to ensure that the two modes intersect at a small angle, and an optical fiber is used to scan along the x-axis capturing two-photon probability. Figure 4(b) shows the measured two-photon spatial beating fringe. Although the single counts of $D_1$ and $D_2$ show a smooth distribution, the coincidence counts reveal an interference fringe with the period of 48 μm. Considering the intersection angle of 10.9 mrad estimated from the combination of mirror and prism with a 1064 nm laser, the results consist well with Eq. (7), which confirms the fringe as the two-photon beating effect. The equivalent wavelength is reduced to $\lambda/2$ =532 nm, which is a principle demonstration of quantum lithography by two-photon NOON state. The visibility of interference fringe is 68.2±1.8%, and it reaches 82.4±1.8% when accidental coincidence counts are excluded from the raw data, therefore the two-photon modes are coherently superposed. The nonideal visibility is caused by the non-perfect mode overlapping as well as the 1:2 intensity-imbalance between the two modes since they go through different optical elements before the interaction. The intensity-imbalance is valued through two-photon coincidence measurement for each mode by blocking the other one. Furthermore, to examine the two-photon coherence length which is theoretically determined by the pump, we record the two-photon spatial beating fringe when the path difference between two modes is far beyond the single photon coherence length. In Fig. 4(c), when inserting a glass plate with high transmission coating for 1064 nm to extend one of the optical path lengths by about 1 mm, which is much larger than the single photon coherence length of 113 μm (estimated by the 10 nm-bandwidth of the interference filter), we find that the two-photon beating fringe keeps almost the same visibility as in Fig. 4(b). This experiment clearly verifies that the interference fringes in Figs. 4(b) and 4(c) are attributed to two-photon interference. For each beamlike mode, the conversion efficiency, defined as the probability that one pump photon has to split into an entangled photon pair, is ~$4.8*10^{-10}$ when the power of pump is 14 mW.

The spatial form of the down converted beams from the crystal is controllable and versatile, which plays the key role in the integrated generation of photonic path-entangled states. The heralded single-photon and beam-like two-photon path-entanglement can be directly generated and easily switched from a monothetic domain-engineered nonlinear photonic crystal. The crystal supplies as a coherent beam splitter inherently within the SPDC processes to distribute photons into desirable spatial modes. It is worth noting that for two concurrent SPDC processes, the spatial form of photon pair seems to be similar with the one from the noncollinear Type-II SPDC process[43]. Although both exhibit as two conical beams, there exist intrinsic differences. The photon pair from the HPLT belong to the same conical beam or specifically the same beamlike spot while for the Type-II case, the signal and idler photons usually emit into different cones. Most importantly, the noncollinear beamlike emission of two-photon NOON state presents unique advantages like easy collection and high-degree of path-entanglement since $|1,1\rangle$ term in this sample is prohibited by the quasi-phase-matching rules.

Considering the state-of-the-art domain-engineered technique, the compact photonic path-entangled state can be extended into high-dimensional fashion and also the multi-photon level, approaching the high-dimensional mode-entanglement or high-number NOON state[18]. Additionally, other degrees of freedom like polarization, orbital angular momentum and frequency can also be manipulated inside the same crystal chip, which will result in a compact generation of other new types of photon-entanglement including hyper-entanglement over several degrees of freedom. The integrated and miniaturized quantum light source based on concurrent multiple QPM processes will act as a key element in exploring the knowledge boundaries of quantum mechanics and prompting the developments of practical quantum technologies.

The authors thank Z. Y. Ou for helpful discussions. This work was supported by the State Key Program for Basic Research in China (nos. 2012CB921802 and 2011CBA00205), the National Natural Science Foundations of China (contract nos. 91121001, 11174121, 11021403 and 11004096), and the Project Funded by the Priority Academic Program development of Jiangsu Higher Education Institutions (PAPD).